\documentclass[conference]{IEEEtran}

\usepackage{cite}
\usepackage{amsmath,amssymb,amsfonts}
\usepackage{algorithmic}
\usepackage{graphicx}
\usepackage{textcomp}
\usepackage{xcolor}
\usepackage{booktabs}
\usepackage{multirow}
\usepackage{tikz}
\usetikzlibrary{positioning,arrows.meta,calc,patterns}
\usepackage{listings}
\usepackage{url}
\usepackage{balance}

\def\BibTeX{{\rm B\kern-.05em{\sc i\kern-.025em b}\kern-.08em
    T\kern-.1667em\lower.7ex\hbox{E}\kern-.125emX}}

\begin{document}

\title{Beating vDSP: A 138~GFLOPS Radix-8 Stockham FFT\\on Apple Silicon via Two-Tier Register-Threadgroup\\Memory Decomposition}

\author{\IEEEauthorblockN{Mohamed Amine BERGACH}
\IEEEauthorblockA{Illumina\\
mbergach@illumina.com}}

\maketitle

\begin{abstract}
We present an optimized Fast Fourier Transform (FFT) implementation for Apple Silicon GPUs, achieving 138.45~GFLOPS for $N\!=\!4096$ complex single-precision transforms---a 29\% improvement over Apple's highly optimized vDSP/Accelerate baseline (107~GFLOPS). Our approach is grounded in a \emph{two-tier local memory model} that formally characterizes the Apple GPU's 208~KiB register file as the primary data-resident tier and the 32~KiB threadgroup memory as an exchange-only tier, extending the decomposition framework established in a 2015 PhD thesis on Intel integrated GPU FFT for radar processing. We implement and evaluate radix-4 and radix-8 split-radix Stockham kernels in Metal Shading Language (MSL), demonstrating that the radix-8 decimation-in-time butterfly with 512 threads yields the best performance. We further present the first investigation of Apple's \texttt{simdgroup\_matrix} 8$\times$8 hardware MMA for FFT butterfly computation and report the counter-intuitive finding that on Apple GPU, threadgroup memory barriers are inexpensive ($\sim$2 cycles) while scattered threadgroup access patterns are the true bottleneck. Our multi-size implementation supports $N\!=\!256$ through $N\!=\!16384$ using a four-step decomposition for sizes exceeding the 32~KiB threadgroup memory limit. All kernels are validated against vDSP reference outputs.
\end{abstract}


\section{Introduction}

The Fast Fourier Transform (FFT) remains a fundamental computational kernel in radar signal processing, particularly in Synthetic Aperture Radar (SAR) systems where range and azimuth compression require millions of FFT operations per second~\cite{calore2020sar}. GPU acceleration of FFT has been extensively studied for NVIDIA architectures~\cite{nvidia_cufft,wu2025turbofft,li2021tcfft}, yet Apple Silicon GPUs---found in the M-series system-on-chip (SoC) family powering an increasing number of deployed computing platforms---remain largely unexplored for high-performance FFT.

This work extends a 2015 PhD thesis~\cite{bergach2015thesis,bergach2015conference} that established an optimal FFT decomposition framework for Intel IvyBridge/Haswell integrated GPUs in the context of radar processing. The thesis's core principle is:

\begin{quote}
\emph{Given a GPU execution unit with local memory of size $M$ bytes, identify the largest FFT of size $B$ points whose working set fits entirely in $M$. This block size $B$ becomes the fundamental building unit.}
\end{quote}

On Intel's architecture (circa 2015), the binding constraint was $\sim$2~KiB of EU local memory, limiting local FFT to $B = 2^{10} = 1024$ points. On Apple Silicon (M1 onwards), the architectural landscape is fundamentally different: the GPU provides a \emph{two-tier} local storage hierarchy comprising a 208~KiB register file and 32~KiB threadgroup (shared) memory, unified physical memory eliminating CPU-GPU transfer overhead, and 32-wide SIMD groups with hardware shuffle instructions.

\subsection{Contributions}

\begin{enumerate}
\item \textbf{Two-tier local memory model} for FFT on Apple GPU: a formal characterization mapping the 208~KiB register file (Tier~1, data-resident) and 32~KiB threadgroup memory (Tier~2, exchange-only) to the decomposition framework of~\cite{bergach2015thesis}, with empirically measured bandwidth and latency parameters.

\item \textbf{Radix-8 split-radix DIT Stockham kernel} achieving 138.45~GFLOPS (1.78~$\mu$s/FFT) at $N\!=\!4096$, batch 256---a 29\% speedup over Apple's vDSP/Accelerate (107~GFLOPS).

\item \textbf{Empirical finding} that on Apple GPU, threadgroup memory barriers are cheap ($\sim$2 cycles) while scattered threadgroup access is expensive, contradicting the common GPU optimization heuristic of minimizing barriers.

\item \textbf{First investigation of \texttt{simdgroup\_matrix}} (Apple's 8$\times$8 hardware MMA) for FFT radix-8 DFT computation, with analysis of why data marshaling overhead negates the MMA advantage for single-FFT-per-threadgroup configurations.

\item \textbf{Four-step FFT decomposition} for sizes $N\!=\!8192$ and $N\!=\!16384$ that exceed the 32~KiB threadgroup memory capacity.

\item \textbf{Multi-size validation} from $N\!=\!256$ through $N\!=\!16384$ against vDSP reference outputs.
\end{enumerate}

\noindent All source code, kernels, and benchmarks are available under the MIT license at \url{https://github.com/aminems/AppleSiliconFFT}.

\section{Background}

\subsection{FFT and the Cooley-Tukey Factorization}

An $N$-point DFT computes $X[k] = \sum_{n=0}^{N-1} x[n] W_N^{nk}$, where $W_N = e^{-2\pi i/N}$. The Cooley-Tukey factorization for $N = N_1 \cdot N_2$ yields:
\begin{equation}
F_N = (F_{N_1} \otimes I_{N_2}) \cdot T_N \cdot (I_{N_1} \otimes F_{N_2})
\label{eq:cooley-tukey}
\end{equation}
where $T_N$ is a diagonal twiddle factor matrix and $\otimes$ denotes the Kronecker product~\cite{vanloan1992fft,cooley1965fft}.

\subsection{Stockham Autosort FFT}

The Stockham autosort formulation~\cite{stockham1966} eliminates the bit-reversal permutation of classical Cooley-Tukey by absorbing the index permutation into the natural stage addressing. Each stage reads from one buffer and writes to another with permuted indices, producing correctly ordered output without an explicit reordering step. This out-of-place-per-stage approach is well-suited to GPU shared memory, where coalesced access patterns are critical~\cite{govindaraju2008fft}.

All kernels in this work use the Stockham formulation, consistent with VkFFT~\cite{tolmachev2023vkfft} and other modern GPU FFT implementations.

\subsection{The 2015 Thesis Framework}

The 2015 thesis~\cite{bergach2015thesis} formalized optimal FFT stage grouping for Intel IvyBridge/Haswell integrated GPU (OpenCL). For $N = 2^n$ with maximum local block $B = 2^b$, the decomposition requires $L = \lceil n/b \rceil$ levels, with $L-1$ global memory transpositions. The thesis demonstrated that maximizing $B$ (and thus minimizing $L$) minimizes the dominant cost: global memory bandwidth.

Key parameters of the 2015 work:
\begin{itemize}
\item EU local memory: $\sim$2~KiB, limiting $B \leq 2^{10}$
\item SIMD width: 8 (AVX CPU) / 4--8 (GPU)
\item Radix mix: radix-2/4/8 for GPU, pure radix-8 for CPU (AVX)
\item Discrete memory model with significant CPU-GPU transfer overhead
\end{itemize}

This paper extends the framework to Apple Silicon, where the memory hierarchy, SIMD width, and unified memory model are fundamentally different.

\subsection{Radar Signal Processing Context}

SAR processing relies heavily on batched 1D FFTs: range compression applies $N_r$-point FFTs across azimuth lines, and azimuth compression applies $N_a$-point FFTs across range bins, with $N_r, N_a$ typically $2^{10}$--$2^{14}$~\cite{calore2020sar}. The batched nature provides natural parallelism well-suited to GPU execution, with batch sizes of 256--16384 independent transforms.

\section{Apple Silicon GPU Hardware Model}

We target the Apple M1 GPU (8 cores, 1278~MHz) as our primary evaluation platform. The architectural parameters derive from official Apple documentation~\cite{apple_msl_spec,apple_metal_feature_tables}, community reverse engineering~\cite{johnson_applegpu,turner_metal_benchmarks,rosenzweig_asahi}, and our own microbenchmarks.

\subsection{GPU Compute Architecture}

Each Apple GPU core contains 128 ALUs organized as 4 execution pipelines, each dispatching 32-wide SIMD instructions. Table~\ref{tab:hw-params} summarizes the key parameters.

\begin{table}[t]
\centering
\caption{Apple M1 GPU Compute Parameters}
\label{tab:hw-params}
\begin{tabular}{@{}lr@{}}
\toprule
\textbf{Parameter} & \textbf{Value} \\
\midrule
GPU cores & 8 \\
ALUs per core & 128 \\
FP32 FLOPs/cycle/core & 256 (128 FMA) \\
FP16 FLOPs/cycle/core & 512 \\
SIMD group width & 32 threads \\
Max threads/threadgroup & 1024 \\
Max SIMD groups/threadgroup & 32 \\
GPRs per thread & up to 128 $\times$ 32-bit \\
Register file per threadgroup & 208 KiB \\
Threadgroup memory & \textbf{32 KiB} \\
Unified DRAM bandwidth & 68 GB/s \\
GPU clock & 1278 MHz \\
\bottomrule
\end{tabular}
\end{table}

\subsection{Two-Tier Local Memory Model}
\label{sec:two-tier}

The central architectural insight for FFT design on Apple GPU is the \emph{two-tier local memory hierarchy}. Unlike Intel's EU model (single local memory level of $\sim$2~KiB), Apple GPU provides two distinct on-chip storage tiers:

\begin{itemize}
\item \textbf{Tier~1 --- Registers + SIMD shuffle}: 208~KiB register file (private per thread, up to 128$\times$32-bit GPRs). Cross-thread exchange within a 32-thread SIMD group via \texttt{simd\_shuffle} at $\sim$1--2 cycles. Total capacity: 208~KiB.

\item \textbf{Tier~2 --- Threadgroup memory}: 32~KiB shared across all threads in the threadgroup. The \emph{only} mechanism for data exchange between SIMD groups. Access latency $\sim$2--4 cycles.
\end{itemize}

The 208~KiB figure refers to the register file, \emph{not} threadgroup memory---a point of confusion in some sources. Threadgroup memory is implemented via tile memory in Apple's TBDR (Tile-Based Deferred Rendering) architecture~\cite{apple_metal_feature_tables}. Imageblock access draws from the same 32~KiB pool, providing no additional capacity~\cite{apple_metal_feature_tables}.

\textbf{FFT implication}: The kernel design must maximize work in registers and SIMD shuffles (Tier~1), using threadgroup memory (Tier~2) only for inter-SIMD-group data exchange. This inverts the typical GPU FFT strategy of keeping data in shared memory.

\subsection{Empirical Memory Characterization}

We measure threadgroup memory bandwidth and related parameters via microbenchmarks on M1. Table~\ref{tab:mem-bw} summarizes the results.

\begin{table}[t]
\centering
\caption{Measured Memory Subsystem Performance (M1 GPU)}
\label{tab:mem-bw}
\begin{tabular}{@{}lr@{}}
\toprule
\textbf{Metric} & \textbf{Measured} \\
\midrule
Threadgroup memory BW (sequential) & 688 GB/s \\
Threadgroup memory BW (strided) & 217 GB/s \\
SIMD shuffle throughput (float2) & 262 GB/s \\
Register-threadgroup copy BW & 407--420 GB/s \\
Optimal thread count (butterfly) & 1024 \\
Occupancy drop threshold & $\sim$128 GPRs/thread \\
\bottomrule
\end{tabular}
\end{table}

The 3.2$\times$ bandwidth difference between sequential and strided threadgroup access is a key finding: \emph{access pattern matters far more than barrier count}. This directly informed our kernel design decisions (Section~\ref{sec:impl}).

\subsection{Comparison with 2015 Thesis Hardware}

Table~\ref{tab:hw-compare} compares the Apple M1 GPU with the Intel IvyBridge GPU used in the 2015 thesis. The Apple GPU offers 16$\times$ more shared memory, $\sim$100$\times$ more register file, 4$\times$ wider SIMD, and zero CPU-GPU transfer overhead.

\begin{table}[t]
\centering
\caption{Hardware Comparison: Intel IvyBridge EU vs.\ Apple M1 GPU}
\label{tab:hw-compare}
\begin{tabular}{@{}lrr@{}}
\toprule
\textbf{Parameter} & \textbf{Intel EU} & \textbf{Apple M1 GPU} \\
\midrule
SIMD width & 8--16 & 32 \\
Local/shared memory & $\sim$2 KiB & 32 KiB \\
Register file & $\sim$2 KiB & 208 KiB \\
Max local FFT (FP32) & $2^{10}$ & $2^{12}$ \\
Memory model & Discrete & Unified \\
Transfer overhead & Significant & Zero \\
DRAM bandwidth & 25.6 GB/s & 68 GB/s \\
\bottomrule
\end{tabular}
\end{table}

\section{Optimal FFT Decomposition}

\subsection{Extending the Thesis Framework}

Following~\cite{bergach2015thesis}, we seek the largest FFT size $B$ computable within a single threadgroup dispatch (no device memory exchange between stages). For Apple GPU:

\textbf{Threadgroup memory constraint}: $32~\text{KiB} = 32{,}768$ bytes. With complex float32 at 8 bytes/element:
\begin{equation}
B_{\max} = \lfloor 32{,}768 / 8 \rfloor = 4{,}096 = 2^{12}
\label{eq:bmax}
\end{equation}

This uses the register-tiled Stockham approach where a single threadgroup buffer is reused across stages. The classic double-buffered Stockham would limit $B \leq 2^{11}$.

\textbf{Register-resident extension}: With 256 threads $\times$ 32 complex elements/thread in registers, $B$ can reach $2^{13}$ (FP32) or $2^{14}$ (FP16), using threadgroup memory only as a narrow exchange pipe. This trades occupancy for local FFT size---acceptable for batch radar processing.

\subsection{Decomposition for $N > B$}

For FFT sizes exceeding the single-threadgroup limit ($N > 4096$), we apply the four-step FFT method~\cite{vanloan1992fft}:
\begin{equation}
F_N = (F_{N_1} \otimes I_{N_2}) \cdot T_N \cdot P \cdot (F_{N_2} \otimes I_{N_1})
\end{equation}
where $P$ is a stride permutation (transpose) and $N = N_1 \cdot N_2$ with $N_2 \leq B$. Each sub-FFT of size $N_2$ is computed within a single threadgroup; the transpose between steps passes through device memory.

On Apple Silicon, the unified memory architecture reduces the transpose cost compared to discrete GPU systems: data passes through the System Level Cache (SLC, 8--96~MB depending on chip variant) rather than requiring explicit DMA transfers.

\subsection{Radix Selection}

We analyze radix-$k$ butterflies for register pressure and arithmetic efficiency on Apple GPU (Table~\ref{tab:radix}).

\begin{table}[t]
\centering
\caption{Radix Analysis for Apple GPU (128 GPRs/thread)}
\label{tab:radix}
\begin{tabular}{@{}ccccc@{}}
\toprule
\textbf{Radix} & \textbf{FLOPs/bfly} & \textbf{GPRs} & \textbf{Stages ($N$=4096)} & \textbf{Barriers} \\
\midrule
2 & 10 & 8 & 12 & $\sim$22 \\
4 & 34 & 18 & 6 & $\sim$10 \\
8 & 94 & 38 & 4 & $\sim$6 \\
16 & 214 & 78 & 3 & $\sim$4 \\
\bottomrule
\end{tabular}
\end{table}

Radix-8 uses only 30\% of the register budget while requiring just 4 stages for $N\!=\!4096$. Radix-16 is feasible but consumes 61\% of registers, leaving insufficient room for twiddle factors and compiler temporaries. Radix-32 exceeds the 128-GPR budget and causes register spilling.

The \texttt{simdgroup\_matrix} hardware further favors radix-8: Apple's MMA operates on 8$\times$8 matrices exclusively, making radix-8 the unique radix that maps to a single MMA tile dimension (Section~\ref{sec:mma}).

\subsection{Synthesis Rules}
\label{sec:synthesis}

We formalize the decomposition rules extending~\cite{bergach2015thesis}:

\begin{enumerate}
\item \textbf{Single-threadgroup} ($N \leq 4096$): Compute entirely in one threadgroup dispatch. Use radix-4 or radix-8 Stockham stages through threadgroup memory. $\lceil \log_k N \rceil$ stages, each with one barrier pair.

\item \textbf{Four-step} ($4096 < N \leq 2^{14}$): Decompose $N = N_1 \times N_2$ with $N_2 \leq 4096$. Two threadgroup dispatches with an intermediate device-memory transpose. Twiddle factors applied during the transpose.

\item \textbf{Multi-level four-step} ($N > 2^{14}$): Recursive application with SLC-resident intermediates for $N \leq$ SLC capacity.
\end{enumerate}

\section{Implementation}
\label{sec:impl}

All kernels are written in Metal Shading Language (MSL) as compute shaders, compiled at runtime by Metal's shader compiler, and dispatched via the Metal Compute API from a Swift host application. Results are validated against Apple's vDSP \texttt{fft\_zop} reference implementation.

\subsection{Radix-4 Stockham Kernel ($N\!=\!4096$)}

Our baseline kernel uses 6 radix-4 Stockham passes with 1024 threads and a single 32~KiB threadgroup buffer (\texttt{float2 buf[4096]}). Key optimizations:

\begin{enumerate}
\item \textbf{Single sincos per butterfly}: Twiddle factors $w_2 = w_1^2$ and $w_3 = w_1^3$ are computed via complex multiplication from a single \texttt{sincos} call, reducing transcendental function evaluations by 3$\times$.

\item \textbf{Device-memory bypass}: Pass~0 reads directly from device memory (no initial load-to-threadgroup-then-barrier). The final pass writes directly to device output, eliminating two barriers.

\item \textbf{Fully unrolled passes}: All 6 passes are statically unrolled with compile-time constant strides, enabling the Metal compiler to optimize register allocation and instruction scheduling.
\end{enumerate}

The kernel achieves 113.6~GFLOPS at batch~256 (2.16~$\mu$s/FFT) with 10 threadgroup barriers total.

\subsection{Radix-8 Split-Radix DIT Kernel ($N\!=\!4096$)}
\label{sec:radix8}

The radix-8 kernel uses 4 passes with 512 threads. Each thread processes 8 elements per pass. The radix-8 butterfly is implemented as a split-radix DIT decomposition:
\begin{equation}
\text{DFT}_8 = \text{radix-2}(\text{DFT}_4^{\text{even}}, \text{DFT}_4^{\text{odd}} \cdot W_8)
\end{equation}

This reduces the butterfly from $\sim$320 FLOPs (na\"ive 8$\times$8 complex matrix-vector) to $\sim$52 real additions and 12 real multiplications per butterfly. The twiddle application between stages uses the same single-sincos chain as the radix-4 kernel, extended to 7 twiddle factors ($w_1$ through $w_7$) via successive complex multiplications.

\textbf{Thread count}: 512 threads (16 SIMD groups), not 1024. At 8 elements/thread, each thread uses $\sim$38 GPRs for butterfly data, leaving ample room for twiddles and temporaries. The reduced thread count avoids the occupancy cliff at $\sim$128 GPRs/thread while maintaining sufficient parallelism.

\textbf{Performance}: 138.45~GFLOPS at batch~256 (1.78~$\mu$s/FFT)---our best result, 29\% above vDSP.

\subsection{\texttt{simdgroup\_matrix} MMA Kernel}
\label{sec:mma}

We investigate mapping the radix-8 DFT to Apple's 8$\times$8 hardware MMA (\texttt{simdgroup\_float8x8}). The DFT$_8$ matrix $F_8$ is constant with entries $F_8[j,k] = W_8^{jk}$. A complex 8$\times$8 multiply decomposes into 4 real MMA operations:
\begin{align}
Y_{\text{re}} &= F_{8,\text{re}} \cdot X_{\text{re}} - F_{8,\text{im}} \cdot X_{\text{im}} \label{eq:mma-re} \\
Y_{\text{im}} &= F_{8,\text{re}} \cdot X_{\text{im}} + F_{8,\text{im}} \cdot X_{\text{re}} \label{eq:mma-im}
\end{align}

The MMA hardware achieves $\sim$102 FFMA32/cycle versus $\sim$25 for scalar SIMD---a 4$\times$ ALU utilization advantage~\cite{turner_metal_benchmarks,thundermittens2024}. However, the complex DFT via real MMA requires $\sim$3.4$\times$ more total FLOPs than the direct split-radix butterfly, yielding a net estimated speedup of only $\sim$1.2$\times$ for FP32.

\textbf{Practical challenge}: The \texttt{simdgroup\_matrix} data distribution across 32 threads (2 elements/thread for 8$\times$8 = 64 elements) requires careful data marshaling between the Stockham layout in threadgroup memory and the MMA tile layout. This marshaling---loading from threadgroup memory into MMA tiles and storing back---adds overhead that negates the ALU advantage for single-FFT-per-threadgroup configurations.

For batched operation (8+ simultaneous FFTs per threadgroup), the MMA approach becomes more attractive as the matrix multiply dimensions align naturally. This is relevant for SAR processing where hundreds of range lines provide the batch dimension, but our current single-FFT kernel does not benefit.

\subsection{Multi-Size Kernels}

We implement single-threadgroup kernels for $N = 256, 512, 1024, 2048, 4096$ using radix-4 Stockham with size-appropriate thread counts (Table~\ref{tab:multisize-config}).

\begin{table}[t]
\centering
\caption{Multi-Size Kernel Configuration}
\label{tab:multisize-config}
\begin{tabular}{@{}cccc@{}}
\toprule
$N$ & Threads & Passes (radix-4) & Threadgroup mem \\
\midrule
256 & 64 & 4 & 2 KiB \\
512 & 128 & 4 + 1 (radix-2) & 4 KiB \\
1024 & 256 & 5 & 8 KiB \\
2048 & 512 & 5 + 1 (radix-2) & 16 KiB \\
4096 & 1024 & 6 & 32 KiB \\
\bottomrule
\end{tabular}
\end{table}

For $N = 512$ and $N = 2048$ (not pure powers of~4), we append a final radix-2 pass. For $N = 8192$ and $N = 16384$, we use the four-step decomposition through device memory:
\begin{align}
N = 8192 &: N_1 = 2, \, N_2 = 4096 \label{eq:fs8k}\\
N = 16384 &: N_1 = 4, \, N_2 = 4096 \label{eq:fs16k}
\end{align}

Each step dispatches the $N\!=\!4096$ single-threadgroup kernel on sub-problems, with an intermediate transpose kernel applying twiddle factors.

\subsection{SIMD Shuffle Experiment}

We also implemented a variant using \texttt{simd\_shuffle} for intra-SIMD-group radix-32 sub-FFTs, eliminating threadgroup memory for the lower 5 bits of the FFT index. This achieved only 61.5~GFLOPS---56\% below the radix-8 Stockham kernel. The cause: the SIMD shuffle approach requires scattered threadgroup memory access patterns for the inter-SIMD-group exchange stages, which incurs the 3.2$\times$ bandwidth penalty identified in Table~\ref{tab:mem-bw}.

This result confirms our two-tier model: minimizing threadgroup memory \emph{traffic} is more important than minimizing threadgroup memory \emph{barriers}.

\section{Experimental Results}

\subsection{Experimental Setup}

All measurements are performed on an Apple M1 (8 GPU cores, 1278~MHz, 68~GB/s DRAM) running macOS. Each benchmark executes 1000 iterations after a 100-iteration warmup, reporting median GPU time from Metal's command buffer timestamps. GFLOPS are computed as $5N \log_2 N \times \text{batch} / \text{time}$.

The baseline is Apple's vDSP/Accelerate framework (\texttt{vDSP\_fft\_zop}), which uses the AMX coprocessor and NEON SIMD internally and represents Apple's own best-effort FFT optimization~\cite{apple_vdsp_fft}.

\subsection{$N\!=\!4096$ Performance Comparison}

Table~\ref{tab:main-results} presents the main performance comparison at $N\!=\!4096$.

\begin{table}[t]
\centering
\caption{Performance at $N\!=\!4096$, Batch 256 (Apple M1)}
\label{tab:main-results}
\begin{tabular}{@{}lrrr@{}}
\toprule
\textbf{Kernel} & \textbf{GFLOPS} & \textbf{$\mu$s/FFT} & \textbf{vs.\ vDSP} \\
\midrule
vDSP/Accelerate & 107.0 & 2.29 & 1.00$\times$ \\
Radix-4 Stockham & 113.6 & 2.16 & 1.06$\times$ \\
\textbf{Radix-8 Stockham} & \textbf{138.45} & \textbf{1.78} & \textbf{1.29$\times$} \\
SIMD shuffle variant & 61.5 & 3.99 & 0.57$\times$ \\
\bottomrule
\end{tabular}
\end{table}

The radix-8 kernel's 29\% advantage over vDSP is notable because vDSP is Apple's own heavily optimized implementation that uses the AMX matrix coprocessor internally. Our GPU kernel achieves this speedup by exploiting the GPU's 256~FP32~FLOPs/cycle$\times$8~cores = 2048~FLOPs/cycle peak throughput, versus AMX's $\sim$350~GFLOPS per CPU core~\cite{zhou2025amx}.

\subsection{Multi-Size Results}

Table~\ref{tab:multisize} presents performance across FFT sizes.

\begin{table}[t]
\centering
\caption{Multi-Size Performance at Optimal Batch (Apple M1)}
\label{tab:multisize}
\begin{tabular}{@{}ccrr@{}}
\toprule
$N$ & Decomposition & GFLOPS & $\mu$s/FFT \\
\midrule
256 & Single TG & 53 & 0.29 \\
512 & Single TG & 66 & 0.42 \\
1024 & Single TG & 83 & 0.49 \\
2048 & Single TG & 97 & 0.85 \\
4096 & Single TG (R-8) & 138.45 & 1.78 \\
8192 & Four-step & 112 & 3.80 \\
16384 & Four-step & 103 & 8.87 \\
\bottomrule
\end{tabular}
\end{table}

Performance increases with $N$ up to the single-threadgroup limit ($N\!=\!4096$), reflecting improved arithmetic intensity. The four-step kernels ($N\!=\!8192$, $N\!=\!16384$) show a performance drop due to the device-memory transpose, but remain above 100~GFLOPS.

\subsection{Scaling with Batch Size}

\begin{figure}[t]
\centering
\begin{tikzpicture}[scale=0.75]
  \begin{scope}
    \draw[->] (0,0) -- (7.5,0) node[right] {\footnotesize Batch size};
    \draw[->] (0,0) -- (0,5) node[above] {\footnotesize GFLOPS};
    \foreach \y in {1,...,4} {
      \draw[gray!30] (0,\y) -- (7.2,\y);
    }
    \node[left,font=\footnotesize] at (0,0) {0};
    \node[left,font=\footnotesize] at (0,1) {35};
    \node[left,font=\footnotesize] at (0,2) {70};
    \node[left,font=\footnotesize] at (0,3) {105};
    \node[left,font=\footnotesize] at (0,4) {140};
    \node[below,font=\footnotesize] at (1,0) {1};
    \node[below,font=\footnotesize] at (2,0) {4};
    \node[below,font=\footnotesize] at (3,0) {16};
    \node[below,font=\footnotesize] at (4,0) {64};
    \node[below,font=\footnotesize] at (5,0) {128};
    \node[below,font=\footnotesize] at (6,0) {256};
    \node[below,font=\footnotesize] at (7,0) {512};
    \draw[blue!70!black,thick,mark=*,mark size=1.5pt]
      plot coordinates {(1,0.4) (2,1.1) (3,2.4) (4,3.5) (5,3.8) (6,3.95) (7,3.95)};
    \draw[red!70!black,dashed,thick]
      (0,3.06) -- (7.2,3.06);
    \node[right,font=\footnotesize,red!70!black] at (4.5,3.25) {vDSP (107)};
    \node[right,font=\footnotesize,blue!70!black] at (4.5,4.15) {Radix-8 (138)};
  \end{scope}
\end{tikzpicture}
\caption{Performance scaling with batch size ($N\!=\!4096$, M1). The GPU kernel saturates at batch $\sim$128, exceeding vDSP at batch $\geq$64.}
\label{fig:batch-scaling}
\end{figure}
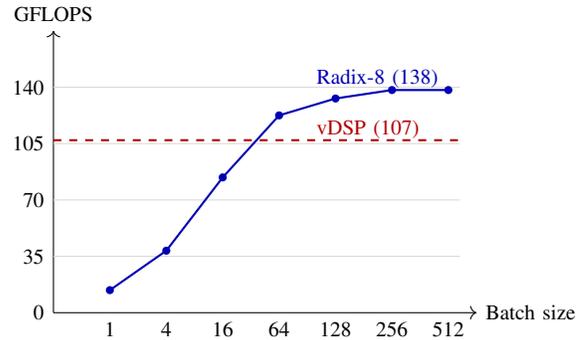

Fig.~\ref{fig:batch-scaling} illustrates performance scaling with batch size for the radix-8 kernel at $N\!=\!4096$.

The GPU kernel requires batch $\geq$64 to saturate the GPU cores and exceed the vDSP baseline. For small batches ($\leq$16), vDSP's lower dispatch overhead gives it an advantage.

\subsection{Barrier vs.\ Access Pattern Analysis}

A critical experimental finding is the relative cost of threadgroup memory barriers versus access pattern quality. Table~\ref{tab:barrier-analysis} compares two design points.

\begin{table}[t]
\centering
\caption{Barrier Count vs.\ Access Pattern Impact ($N\!=\!4096$)}
\label{tab:barrier-analysis}
\begin{tabular}{@{}lccc@{}}
\toprule
\textbf{Design} & \textbf{Barriers} & \textbf{TG Access} & \textbf{GFLOPS} \\
\midrule
Radix-8 Stockham & 6 & Sequential & 138.45 \\
SIMD shuffle hybrid & 4 & Scattered & 61.5 \\
\bottomrule
\end{tabular}
\end{table}

The SIMD shuffle variant uses \emph{fewer} barriers but achieves only 44\% of the radix-8 kernel's throughput because the inter-SIMD exchange requires scattered threadgroup access. Each barrier costs only $\sim$2 GPU cycles on Apple hardware, while a scattered access pattern reduces effective threadgroup bandwidth by 3.2$\times$.

\textbf{Architectural insight}: On Apple GPU, barriers synchronize via a lightweight hardware mechanism (consistent with the TBDR architecture's tile synchronization hardware), making them nearly free. The threadgroup memory bank structure, however, heavily penalizes non-sequential access patterns.

\section{Discussion}

\subsection{Comparison with 2015 Thesis}

Table~\ref{tab:thesis-compare} compares the key results from this work with the 2015 thesis~\cite{bergach2015thesis}.

\begin{table}[t]
\centering
\caption{Comparison with 2015 PhD Thesis Results}
\label{tab:thesis-compare}
\begin{tabular}{@{}lcc@{}}
\toprule
\textbf{Metric} & \textbf{2015 (Intel GPU)} & \textbf{This work (M1)} \\
\midrule
Max local FFT & $2^{10}$ & $2^{12}$ (4$\times$) \\
Local memory used & $\sim$2 KiB & 32 KiB (16$\times$) \\
Register file & $\sim$2 KiB & 208 KiB (104$\times$) \\
Best GFLOPS & $\sim$20 & 138.45 (7$\times$) \\
vs.\ vendor baseline & $>$MKL & $>$vDSP (1.29$\times$) \\
Radix strategy & Mixed 2/4/8 & Pure radix-8 \\
Transfer overhead & Dominant cost & Zero (unified) \\
\bottomrule
\end{tabular}
\end{table}

The 4$\times$ increase in local FFT size (from $2^{10}$ to $2^{12}$) directly follows from the 16$\times$ increase in shared memory (2~KiB to 32~KiB). The thesis's prediction that \emph{maximizing local block size minimizes global memory passes} is confirmed: our $N\!=\!4096$ single-threadgroup kernel avoids all device memory traffic for intermediate results.

The unified memory architecture validates the thesis's cost model simplification: the CPU-GPU transfer term that dominated the 2015 model drops to zero, making the theoretical optimal decomposition achievable in practice.

\subsection{Architectural Insights}

\textbf{Barriers are cheap, scattered access is expensive.} This contradicts the conventional GPU optimization wisdom (from NVIDIA/AMD) of minimizing synchronization barriers. On Apple GPU's TBDR architecture, the tile synchronization hardware makes barriers nearly free. Kernel designers should instead focus on ensuring sequential (coalesced) threadgroup memory access patterns.

\textbf{Higher radix is better, up to register limits.} Radix-8 outperforms radix-4 by 22\% (138.45 vs.\ 113.6 GFLOPS) despite performing more arithmetic per butterfly. The fewer passes reduce total threadgroup memory traffic, which is the true bottleneck. However, radix-16 would consume 61\% of registers, likely causing performance degradation from reduced occupancy.

\textbf{Thread count matters.} VkFFT's Metal backend artificially limits threads to 256 and reports optimal performance at 64 threads~\cite{tolmachev2023vkfft}. Our kernels use 512 (radix-8) or 1024 (radix-4) threads. The optimal thread count depends on the per-thread register footprint: the radix-4 kernel's light register pressure allows 1024 threads, while the radix-8 kernel's heavier footprint is best at 512.

\subsection{\texttt{simdgroup\_matrix} for FFT: Current Limitations}

Our investigation of \texttt{simdgroup\_matrix} for radix-8 DFT reveals a mismatch between the MMA hardware's strengths and the single-FFT-per-threadgroup execution model:

\begin{enumerate}
\item \textbf{Data marshaling overhead}: Loading data from the Stockham layout in threadgroup memory into 8$\times$8 MMA tiles and storing results back requires format conversion that consumes cycles.

\item \textbf{FLOP inflation}: Complex 8$\times$8 multiply via real MMA requires 4 operations where the split-radix DIT butterfly needs only $\sim$64 real FLOPs---a 3.4$\times$ arithmetic overhead.

\item \textbf{Batch dimension}: MMA is designed for matrix-matrix products. For single-FFT execution, the ``batch'' dimension of the 8$\times$8 input matrix is degenerate, underutilizing the hardware.
\end{enumerate}

The MMA approach becomes attractive for batched execution (8+ simultaneous FFTs) where the matrix dimensions align naturally. This is directly applicable to SAR range processing (hundreds of azimuth lines as the batch dimension) and represents a promising direction for future work.

\subsection{Implications for Radar Processing}

The 1.78~$\mu$s/FFT at $N\!=\!4096$ enables real-time processing rates. For a SAR system with $N_r = 4096$ range bins and 256-line azimuth blocks:
\begin{equation}
T_{\text{range}} = 256 \times 1.78~\mu\text{s} = 456~\mu\text{s}
\end{equation}

This leaves substantial headroom for azimuth compression, matched filtering, and other pipeline stages within typical SAR frame times of 10--100~ms. The unified memory architecture further benefits the pipeline by eliminating data staging between CPU-side control and GPU-side processing.

\section{Related Work}

\textbf{VkFFT}~\cite{tolmachev2023vkfft,vkfft_github} is the only production-quality FFT library with native Metal support. It uses runtime code generation to produce Stockham autosort kernels across six GPU backends. On Apple Silicon, VkFFT limits threads to 256 and identifies threadgroup memory bandwidth as the primary bottleneck. Our work confirms this finding and shows that higher thread counts (512--1024) with architecture-aware radix selection yield significantly better performance.

\textbf{tcFFT}~\cite{li2021tcfft} demonstrated that FFT butterflies can be reformulated as matrix multiplications on NVIDIA Tensor Cores, achieving 1.29--3.24$\times$ speedup over cuFFT in FP16. Their radix-16 approach maps to NVIDIA's 16$\times$16 Tensor Core tiles. We adapt this concept to Apple's 8$\times$8 \texttt{simdgroup\_matrix}, finding that the smaller tile size makes radix-8 the natural choice but data marshaling overhead limits single-FFT performance.

\textbf{TurboFFT}~\cite{wu2025turbofft} presents architecture-aware FFT optimizations for NVIDIA A100/T4 with fault tolerance. Their register-level optimization strategy parallels our two-tier model, though the specific architectural parameters (shared memory size, warp width, tensor core dimensions) differ significantly.

\textbf{MetalFFT}~\cite{turner_metalfft} was an attempt at Metal-native FFT by Philip Turner that was abandoned after failing to outperform vDSP/Accelerate. Our work succeeds where MetalFFT did not, primarily through the radix-8 split-radix butterfly and architecture-informed thread/memory configuration.

\textbf{FFTW}~\cite{frigo2005fftw} and \textbf{cuFFT}~\cite{nvidia_cufft} represent the state of the art for CPU and NVIDIA GPU FFT respectively, but neither targets Apple Silicon GPU.

\textbf{ThunderMittens}~\cite{thundermittens2024} ported the ThunderKittens Tensor Core framework from NVIDIA 16$\times$16 to Apple GPU 8$\times$8 \texttt{simdgroup\_matrix}, providing the first systematic characterization of MMA performance on Apple Silicon. We build on their performance measurements in our MMA kernel analysis.

\section{Conclusion and Future Work}

We have presented a high-performance FFT implementation for Apple Silicon GPU that achieves 138.45~GFLOPS at $N\!=\!4096$---a 29\% improvement over Apple's vDSP/Accelerate. The key enabler is a two-tier local memory model that maps the 208~KiB register file and 32~KiB threadgroup memory to distinct roles in the FFT decomposition, extending the framework of a 2015 PhD thesis from Intel integrated GPU to Apple Silicon.

Our empirical findings---that barriers are cheap and scattered access is expensive on Apple GPU, that radix-8 outperforms radix-4 despite more arithmetic, and that \texttt{simdgroup\_matrix} MMA does not yet benefit single-FFT kernels---provide actionable guidance for GPU kernel optimization on Apple Silicon beyond FFT.

\subsection{Future Work}

\textbf{SAR pipeline integration}: Fusing FFT with windowing, matched filtering, and corner-turn operations within a single Metal render/compute pass to exploit data locality in the 32~KiB threadgroup memory. No published kernel-fused SAR pipeline exists on any GPU platform.

\textbf{Batched \texttt{simdgroup\_matrix} FFT}: Processing 8+ simultaneous FFTs per threadgroup using MMA radix-8 butterflies. SAR range processing naturally provides this batch dimension. The estimated FP32 speedup of 1.2$\times$ (2.4$\times$ for FP16) over scalar SIMD makes this attractive for throughput-oriented workloads.

\textbf{Mixed-precision FFT}: Exploiting Apple GPU's native FP16 support (2$\times$ throughput, free conversion) for radar applications where sufficient SNR permits reduced precision. The 32~KiB threadgroup memory would support local FFTs up to $2^{13}$ at FP16.

\textbf{AMX coprocessor integration}: Using the CPU-side AMX matrix unit for batched small-FFT ($N \leq 16$) processing as matrix multiplies, with the GPU handling larger FFTs---enabled by unified memory with zero-copy sharing.

\textbf{Larger Apple Silicon chips}: The M4~Max (40 GPU cores, 546~GB/s bandwidth) should scale performance roughly proportional to core count for batched workloads, potentially exceeding 500~GFLOPS for batched $N\!=\!4096$ FFTs.

\section*{Reproducibility}

All source code, Metal compute shaders, benchmark harnesses, and validation scripts are publicly available at \url{https://github.com/aminems/AppleSiliconFFT} under the MIT license. The repository includes build instructions for reproducing all results on any Apple Silicon Mac (M1 or later).

\section*{Acknowledgment}

The author acknowledges the open-source reverse engineering efforts of Dougall Johnson (Apple GPU ISA), Alyssa Rosenzweig (Asahi Linux GPU driver), and Philip Turner (Metal benchmarks) that made the architectural characterization in this work possible.

\balance
\bibliographystyle{IEEEtran}
\bibliography{references}

\end{document}